\documentclass{Interspeech2024}
\usepackage{xcolor, soul, stfloats}
\usepackage{amsmath}
\usepackage{subfigure}

\interspeechcameraready 

\title{Unsupervised Improved MVDR Beamforming for Sound Enhancement}

\name[affiliation={1}]{Jacob}{Kealey}
\name[affiliation={2}]{John R.}{Hershey}
\name[affiliation={1}]{François}{Grondin}

\address{
  $^1$Université de Sherbrooke, Sherbrooke, QC, Canada\\
  $^2$Google Research, Cambridge, MA, USA }
\email{jacob.kealey@usherbrooke.ca, johnhershey@google.com, francois.grondin2@usherbrooke.ca}


\begin{document}

%Tools
\include{widebar}
\newcommand{\comment}[1]{}

\maketitle

\begin{abstract}
    Neural networks have recently become the dominant approach to sound separation. Their good performance relies on large datasets of isolated recordings. For speech and music, isolated single channel data are readily available; however, the same does not hold in the multi-channel case, and with most other sound classes. Multi-channel methods have the potential to outperform single channel approaches as they can exploit both spatial and spectral features, but the lack of training data remains a challenge. We propose unsupervised improved minimum variation distortionless response (UIMVDR), which enables multi-channel separation to leverage in-the-wild single-channel data through unsupervised training and beamforming. Results show that UIMVDR generalizes well and improves separation performance compared to supervised models, particularly in cases with limited supervised data. By using data available online, it also reduces the effort required to gather data for multi-channel approaches.
\end{abstract}

\section{Introduction}

Humans and animals monitor their environment, detect threats, and communicate using their hearing abilities. Likewise, robots need to able to process both speech and other sounds in their environment in order to interact naturally with the world. Most animals are limited to binaural hearing, but robots can be equipped with more than two microphones. Robot audition entails capturing audio signals with a microphone array to recognize individual signals of interest. In current approaches, deep neural network models are used in sound event detection (SED) \cite{espinosa_click-event_2021, grondin2019sound}, inferring when a particular sound has happened, in sound source localization (SSL) \cite{rascon_localization_2017, grondin2019svd} to determine the \emph{direction of arrival} (DOA) of the sound, in \emph{sound classification} (SC) \cite{xue_sound-based_2022} to infer the class of  the sound, or in \emph{speech recognition} (SR) \cite{bingol_performing_2020} to infer the transcript of speech sounds.

In noisy and reverberant environments, SED \cite{heittola_sound_2011}, SSL \cite{manamperi_drone_2022, grondin2019lightweight}, SC \cite{denton_improving_2022} and SR \cite{xiao_deep_2016} performance degrades because the target signal is mixed with interfering signals. This is especially challenging for robots because their actuators may generate noise and they may interact in noisy and reverberant indoor environments.  To alleviate this, sound separation or sound enhancement can be used prior  to recognition to estimate isolated sounds for use in SED, SSL, SC and SR. Recently, deep learning algorithms have achieved greatly improved sound source separation performance, leading to improvement in downstream recognition tasks such as SED \cite{kong_joint_2018}, SSL \cite{manamperi_drone_2022}, SC \cite{denton_improving_2022} and SR \cite{xiao_deep_2016}.
% Final formating
\newline

Furthermore, it has been shown that using multi-channel input can improve the performance of sound separation and sound enhancement, especially in noisy and reverberant environments \cite{wang_multi-channel_2018}. Single channel approaches are limited to spectral features, whereas multi-channel approaches using microphone arrays can utilize both spatial and spectral features. However, collecting multi-channel data for deep learning approaches can be challenging.

Multi-channel approaches often use datasets that are synthetically created using single channel data and \emph{room impulse responses} (RIRs), in order to provide isolated ground truth sources for supervised training. However, the generated data can differ from real data because of the challenges associated with accurately simulating the reverberation characteristics of an actual room. Custom datasets for a specific microphone array can also be created, but this is time consuming, and models trained on such data generalize poorly to other microphone arrays. The use of beamformers using signal estimates from single channel approaches can overcome these challenges \cite{erdogan_improved_2016, heymann_neural_2016}. Initially, single channel approaches can extract a mask of the signal of interest using spectral features with a reference channel. This mask can then be applied to each channel of the multi-channel input to obtain an estimate of the signal of interest. Subsequently, this estimated signal can be refined using \emph{minimum variance distortionless response} (MVDR) beamforming \cite{souden_optimal_2010}, which uses the spatial features to enhance the target signal \cite{erdogan_improved_2016, heymann_neural_2016, wang2021sequential}. The deep learning algorithm used in this approach can be trained on single-channel input and therefore eliminates the need of multi-channel datasets for processing input from multiple microphones.

Approaches using supervised learning need to know the ground truth to train the deep learning algorithm. In practice, it is infeasible to record both the ground truth signal and the mixture at the same time, without introducing cross-talk between the recordings. To address this problem, unsupervised approaches can be used without the ground truth signals \cite{tzinis_unsupervised_2019, wisdom_unsupervised_2020, wisdom_sparse_2021, han2023unsupervised}. This allows the use of databases of real life recordings, also known as in-the-wild data, to train deep learning algorithms.

\emph{Unsupervised improved MVDR beamforming} (UIMVDR) combines the multi-channel beamforming approach with the single-channel unsupervised approach to enhance a sound of interest. This enables multi-channel sound enhancement to benefit from large real single channel databases like recordings made on phones or recordings available on video sharing services. We also propose a new dataset to evaluate supervised, unsupervised, single-channel, and multi-channel sound separation algorithms. The code and evaluation dataset are available online \footnote{https://github.com/introlab/uimvdr}.

\begin{figure*}[t]
  \centering
  \includegraphics[width=\textwidth]{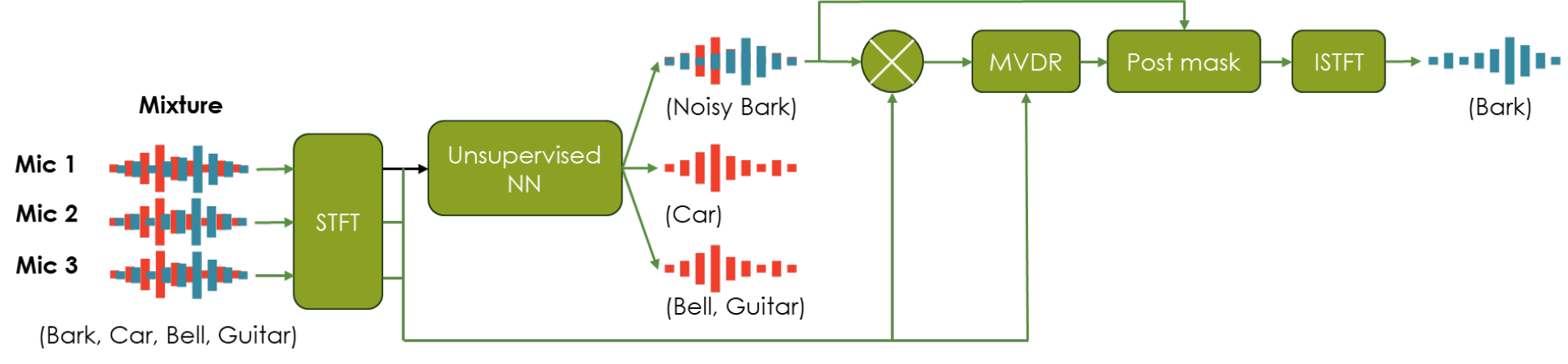}
  \caption{Pipeline of UIMVDR during inference.}
  \label{fig:pipeline}
\end{figure*}

\section{Proposed Method}
The proposed method is shown in figure \ref{fig:pipeline}. The mixture is first encoded using a \emph{short-time Fourier transform} (STFT). The STFT is used as the encoder for compatibility with the MVDR beamformer and because it performs well for sound separation as seen in Kavelerov et al. \cite{kavalerov_universal_2019}. They also report that, for windows longer than 5 ms, STFT outperforms learned encoders. In order to perform well, beamforming needs a window size that is on the order of the reverberation length \cite{wang_sequential_2021}. Window sizes often used are 64 \cite{liu_neural_2018} and 32 ms \cite{zhang_adl-mvdr_2021}. STFT encoders also tend to perform better in reverberant environments \cite{heitkaemper_demystifying_2020}. A single-channel unsupervised neural network is then used to separate the sources. Beamforming is applied using the estimated target signal to improve the enhancement. Finally, the \emph{inverse short-time Fourier Transform (ISTFT)} is used to decode the enhanced signal and obtain the final prediction in the time domain. This technique can also be used for multi-source separation by using each estimated source as a beamforming target.

\subsection{Separation}
The mixture \(y\) is encoded in the frequency domain (\(Y\)) using the STFT. The sound enhancement problem can be mathematically defined in the frequency domain, starting with the forward model:
\begin{align}
  Y(t,f) &= X(t,f) + N(t,f),
\end{align}
where \(X\) is the target signal, \(N\) is the interference, \(t\) is the time frame index and, \(f\) is the frequency bin index.
Neural networks have proved to be capable of solving the inverse problem of separating sound sources using masks, because of their capacity to learn complex non-linear mappings \cite{kavalerov_universal_2019}. This can be defined mathematically as follows:
\begin{align}
  \hat{X}(t,f) &= M(t,f) Y(t,f),
\end{align}
where \(M \in [0, 1]\) is the estimated mask by the neural network and \(\hat{X}\) is the estimated signal. The estimated signal is finally decoded with the ISTFT to obtain the enhanced signal in the time domain \(\hat{x}\).

\subsection{Efficient MixIT}
Separation models can be trained using full supervision \cite{yu_permutation_2017} but this remains difficult for general purpose sound enhancement and sound separation as the clean isolated sources and the mixture are rarely available in real recordings, without significant cross talk. Unsupervised training enables us to use noisy recordings made with everyday devices. This increases the amount of data available for training and eases data collection. To train the neural network without supervision, we used the unsupervised framework, \emph{mixture invariant training} (MixIT) \cite{wisdom_unsupervised_2020, wisdom_sparse_2021}. Unsupervised training also helps with generalizing to multiple environments \cite{wisdom_unsupervised_2020}. MixIT combines two or more mixtures to create a \emph{mixture of mixtures} (MoM). The MoM is fed as the input of the separation model. The model then predicts the separated sources. Using a mixing matrix \(A\), the separated sources are assigned to one of the original mixtures. For each possible mixing matrix, the assigned sources are summed to form reconstructed mixtures. The mixing matrix is selected to minimize the loss between the reconstructed mixtures and the original mixtures.
Although this works well, it can be quite computationally expensive to calculate the loss for every possible assignment.
To address this, \cite{wisdom_sparse_2021} proposes an efficient version of MixIT using the least-squares algorithm:
\begin{align}
  \hat{A} &= \mathcal P_\mathbb{B} (\operatorname{argmin}_{A\in\mathbb{R}^{N \times S}}  ||y - A\hat{x}||_2 ^2),
\end{align}
where \(\mathcal P_\mathbb{B}\) is a projection that sets the maximum of each column to 1, and the rest to 0. \(N\) is the number of mixtures used to create the MoM and \(S\) is the number of sources predicted by the model. Once the estimated mixing matrix \(\hat{A}\) is obtained, it is used to reconstruct the mixtures and compute a signal-level loss on them \cite{wisdom_unsupervised_2020}.

In the case of sound enhancement, a weakly-supervised setting is used. The first mixture always contains the target signal class which can be clean or noisy. If the classification information is not available to create the target split, a sound classifier could be used as mentioned in \cite{wisdom_unsupervised_2020}. The second mixture always contains one or more non-target signals. The combination of both creates the MoMs. For sound enhancement, we usually predict one source for the target, and the interference is the difference between the mixture and the prediction. To use MixIT for sound enhancement, three sources need to be predicted. This requirement stems from the need to reconstruct the original mixture in the presence of interference alongside the target. The assignment matrix is constrained such that the first mixture can be reconstructed using the first output only, the first and second output or the first and third output. This forces the target signal to be predicted in the first output. The second mixture is reconstructed using the outputs not used for the target signal. 

\subsection{MVDR Beamforming}
While using a weakly supervised deep learning model achieves good results for sound enhancement and separation, there is room for improvement. In fact, the mask prediction can sometimes omit part of the target signal or produce residual noise. To address this, we use the predicted signals to compute \emph{spatial covariance matrices} (SCMs)
(\({\bf{\Phi}_{\mathbf{\hat{N}\hat{N}}}}\),
\({\bf{\Phi}_{\mathbf{\hat{X}\hat{X}}}}\))
and then beamform in that direction using MVDR, as follows:
\begin{align}
    {\bf{\Phi}_{\mathbf{\hat{X}\hat{X}}}}(f) =
        \frac{1}{T}
        \sum_{t=1}^{T}\mathbf{\hat{X}}(t,f) \mathbf{\hat{X}}(t,f)^{H}, \\
    \mathbf{\hat{N}}(t,f) = (\mathbf{Y}(t,f)-\mathbf{\hat{X}}(t,f)), \\
    {\bf{\Phi}_{\mathbf{\hat{N}\hat{N}}}}(f) =
        \frac{1}{T}
        \sum_{t=1}^{T} \mathbf{\hat{N}}(t,f) \mathbf{\hat{N}}(t,f)^{H}, \\
    \textbf{F}_{\text{MVDR}}(f) = \frac
    {
        {{\bf{\Phi}_{\mathbf{\hat{N}\hat{N}}}^{-1}}(f)
        {\bf{\Phi}_{\mathbf{\hat{X}\hat{X}}}}}(f)
    }
    {\text{Trace}
        ({{{\bf{\Phi}_{\mathbf{\hat{N}\hat{N}}}^{-1}}(f)
        \bf{\Phi}_{\mathbf{\hat{X}\hat{X}}}}(f))}
    }
    \mathbf{u}, \\
    \widebar{X}(t, f) = \textbf{F}_{\text{MVDR}}^{H}(f)\mathbf{Y}(t, f),
\end{align}
where \(\{...\}^H\) stands for the Hermitian operator and \(\mathbf{u}\) is a one-hot vector indicating the reference microphone. This is possible when multiple channels are available, assuming the sources come from different directions. MVDR beamforming also ensures the linearity of the estimate which is useful when enhancement is applied upstream of another algorithm. To further improve beamforming estimates, a minimum floor post-masking filter is used \cite{erdogan_improved_2016}.
\begingroup

\setlength{\tabcolsep}{7.5pt} % Default value: 6pt
\renewcommand{\arraystretch}{1.25} % Default value: 1
\begin{table*}[t]
  \fontsize{8pt}{8pt}\selectfont
  \caption{SI-SDR improvement (SI-SDRi) for bark enhancement with mixtures containing a target with interference (T+I) and SI-SDR for bark enhancement with mixtures containing the target only (T-Only). Confidence intervals are given using the same method as \cite{wisdom_unsupervised_2020}.}
  \label{tab:results_bark}
  \centering
  \begin{tabular}{ *{11}{c} }
    \toprule
    \multicolumn{3}{c}{}& \multicolumn{2}{c}{\textbf{ReSpeaker}} & \multicolumn{2}{c}{\textbf{Kinect}} &\multicolumn{2}{c}{\textbf{16Sounds}} &\multicolumn{2}{c}{\textbf{Single}} \\
    \cmidrule(lr){4-5}\cmidrule(lr){6-7} \cmidrule(lr){8-9} \cmidrule(lr){10-11}
    \raisebox{\dimexpr1.25\normalbaselineskip-.75\height}[0pt][0pt]{\begin{tabular}{@{}c@{}}
        \textbf{Train Set}
    \end{tabular}}
    & 
    \raisebox{\dimexpr1.25\normalbaselineskip-.75\height}[0pt][0pt]{\begin{tabular}{@{}c@{}}
        \textbf{Method}
    \end{tabular}}
    & 
    \raisebox{\dimexpr1.25\normalbaselineskip-.75\height}[0pt][0pt]{\begin{tabular}{@{}c@{}}
        \textbf{Beamforming}
    \end{tabular}}
    &\textbf{\textit{T+I}} & \textbf{\textit{T-Only}} &
    \textbf{\textit{T+I}} & \textbf{\textit{T-Only}} &
    \textbf{\textit{T+I}} & \textbf{\textit{T-Only}} &
    \textbf{\textit{T+I}} & \textbf{\textit{T-Only}}\\
       &  &  & \textpm0.04 & \textpm0.27 & \textpm0.07 & \textpm0.79 & \textpm0.05 & \textpm0.30 & \textpm0.11 & \textpm4.68 \\
    \midrule
    FSD50K   & Supervised & No & 4.75 & 7.46 & 4.95 & 13.60 & 4.72 & 7.74 & 8.60 & 12.08 \\ % id: 9wb9io4f
             &            & Yes & 9.20 & 13.26 & 8.64 & 20.41 & 11.89 & 20.14 & - & - \\
             & Unsupervised & No & 5.71 & 9.04 & 5.18 & 14.79 & 5.64 & 9.32 & 8.92 & 12.78 \\ % id: 31ypjv7l
             &            & Yes & 10.63 & 14.64 & 9.35 & 20.22 & 13.30 & 20.98 & - & - \\
             & Unsup. w/ Weighting & No & 5.55 & 7.60 & 5.19 & 12.40 & 5.55 & 8.25 & 8.91 & 9.95 \\ % id: hs56ya47
             &            & Yes & 10.66 & 12.84 & 9.60 & 18.07 & 13.67 & 20.33 & - & - \\
    \midrule
    AudioSet & Unsupervised & No & 7.63 & 11.78 & 8.30 & 30.47 & 7.57 & 12.11 & 10.99 & 26.44 \\ % id: bt8rv34b
             &            & Yes & 13.65 & 19.77 & \textbf{13.11} & 38.93 & \textbf{17.09} & \textbf{27.37} & - & - \\
             & Unsup. w/ Weighting & No & 7.65 & 12.07 & 8.25 & 32.90 & 7.64 & 12.04 & \textbf{11.00} & \textbf{31.65} \\ % id: qduqlu78
             &            & Yes & \textbf{13.79} & \textbf{20.01} & 13.10 & \textbf{43.12} & 16.98 & 27.13 & - & - \\
    \bottomrule
    \end{tabular}
\end{table*}

\endgroup

\section{Experiments and discussion}

For our expirements, we use a TDCN++ model \cite{kavalerov_universal_2019} with hyperparameters nearly identical to those used in \cite{wisdom_unsupervised_2020}. They use an instance norm for the normalization layers. Although this gives better results for direct estimation, we found that it gives worse results with beamforming than the original global layer normalization used in Conv-Tasnet \cite{luo_conv-tasnet_2019}. A frame window of 64 ms is used, as it is necessary to have a sufficiently large window for detecting the time differential of arrival of sources between pairs of microphones for beamforming. We use segments of 5 seconds sampled at 16 kHz. The signal-level loss function used is the negative thresholded SNR:
\begin{align}
    \mathcal{L}_{\mathrm{SNR}}(x, \hat{x}) = -10\text{log}_{10}\frac{||x||^2}{||x-\hat{x}||^2 + \tau||y||^2},
\end{align}
where \(\tau = 10^{-\mathrm{SNR}_{\mathrm{max}}/10}\) thresholds the loss at \(\mathrm{SNR}_{\mathrm{max}}\). Wisdom et al. found that \(\mathrm{SNR}_{\mathrm{max}}\) = 30 dB is a good maximum value. In an attempt to reduce the leaking in the target signal, we also propose to minimize the target weight across all frequencies in the loss:
\begin{align}
    \mathcal{L}(x, \hat{x}) = \mathcal{L}_{\mathrm{SNR}}(x, \hat{x}) + \frac{\gamma}{TF}\sum_{t=1}^{T}\sum_{f=1}^{F}|\hat{X}(t,f)|^\beta,
\end{align}
where \(\gamma\) is the weight of the energy loss and \(\beta\) is the exponent that controls the weighting across the frequencies. We use a \(\gamma\) of 0.01 for all train sets and a \(\beta\) of 0.01 for the Freesound Dataset 50K (FSD50K) \cite{fonseca_fsd50k_2022} bark enhancement train set and 0.5 for the remaining train sets.

\begin{figure}
  \centering
  \subfigure[Setup at position A]{\includegraphics[width=0.42\linewidth]{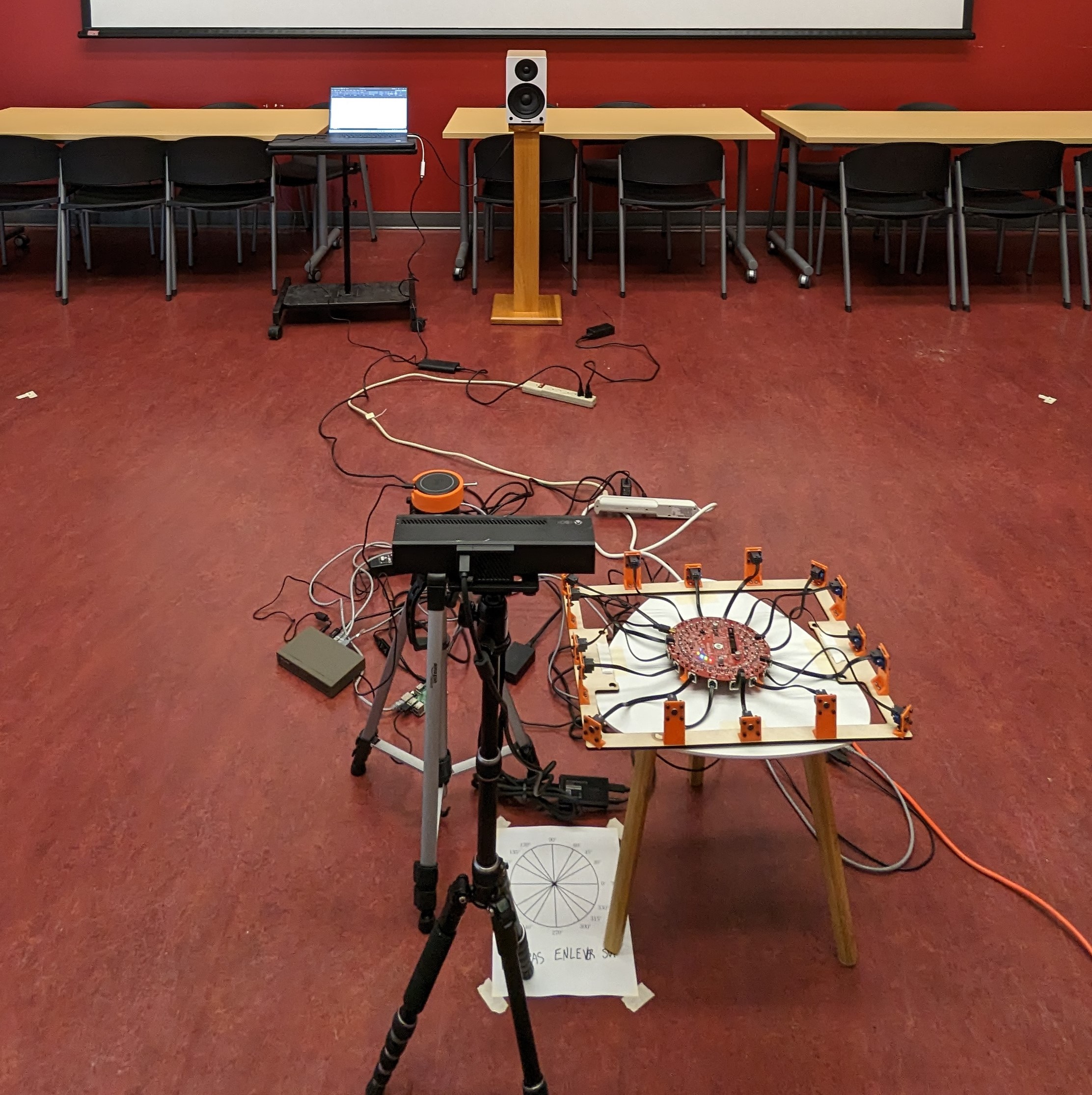}}
  \subfigure[All recording positions.]{\includegraphics[width=0.42\linewidth]{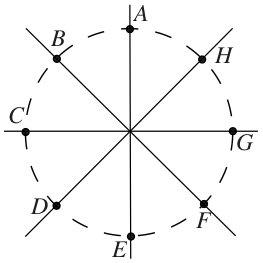}}
  \caption{Recording positions for the MCFSTD.}
  \label{fig:positions_dataset}
\end{figure}

To evaluate UIMVDR, a multichannel dataset with clean sources that comes from different directions is required. Some multichannel datasets are available publicly like STARSS22 \cite{politis_starss22_2022}. However, it does not have the clean sources, whereas SECL-UMons \cite{brousmiche_secl-umons_2020} only has one microphone array and two rooms. This is why we created a custom dataset: the \emph{Multi-Channel Free Sound Test Dataset} (MCFSTD). MCFSTD was recorded on three different microphone arrays in four different rooms. Figure \ref{fig:positions_dataset} a) shows the experimental setup. The first microphone array is a square 4 microphone commercial array, the USB ReSpeaker \footnote{https://wiki.seeedstudio.com/ReSpeaker-USB-Mic-Array/}. The second is a Xbox One Kinect \cite{guzsvinecz2019suitability} which is a 4 microphone linear array. The last microphone array is the 16SoundsUSB from IntRoLab\footnote{https://github.com/introlab/16SoundsUSB} which is a 16 microphone array. The microphones are positioned along the perimeter of two rectangular planes, spaced 3.5 cm apart. The dimensions of the rectangle are 47 cm in length and 36.5 cm in width. As for the rooms, the recordings were made in a conference room, a living room, a dining hall, and a large room used for robotics experiments. 

 As shown in Figure \ref{fig:positions_dataset} b) a loudspeaker \footnote{https://www.fluance.com/powered-2-0-bluetooth-active-5-inch-bookshelf-speakers-bamboo} played a consistent 3-minute audio segment for each of the 10 classes (Bark, Church bell, Coin dropping, Computer keyboard, Mechanical fan, Piano, Printer, Speech, Thunder, Waves) at every 45 degrees on the perimeter of the circle (positions A to G). The arrays were placed at the center of the circle. This means MCFSTD totals 52.8 hours of audio. A chirp was also recorded to compute RIRs if needed. Note that, in the recordings, the loudspeaker introduces a slight distortion in the lower frequencies, the impact of this should be investigated. For the Kinect test dataset, we only use the positions at the front of the matrix for the target, as well as positions C and G, because the microphones are directional.

\begingroup

\setlength{\tabcolsep}{3.75pt} % Default value: 6pt
\renewcommand{\arraystretch}{1.15} % Default value: 1
\begin{table*}[t]
  \fontsize{7.5pt}{7.5pt}\selectfont
  \caption{Results for Speech Enhancement. Confidence intervals are given using the same method as \cite{wisdom_unsupervised_2020}.}
  \label{tab:results_pesq_stoi}
  \centering
  \begin{tabular}{ *{15}{c} }
    \toprule
    \multicolumn{3}{c}{}& \multicolumn{3}{c}{\textbf{ReSpeaker}} & \multicolumn{3}{c}{\textbf{Kinect}} &\multicolumn{3}{c}{\textbf{16Sounds}} &\multicolumn{3}{c}{\textbf{Single}} \\
    \cmidrule(lr){4-6}\cmidrule(lr){7-9} \cmidrule(lr){10-12} \cmidrule(lr){13-15}
    \raisebox{\dimexpr1.25\normalbaselineskip-.75\height}[0pt][0pt]{\begin{tabular}{@{}c@{}}
        \textbf{Train Set}
    \end{tabular}}
    & 
    \raisebox{\dimexpr1.25\normalbaselineskip-.75\height}[0pt][0pt]{\begin{tabular}{@{}c@{}}
        \textbf{Method}
    \end{tabular}}
    & 
    \raisebox{\dimexpr1.25\normalbaselineskip-.75\height}[0pt][0pt]{\begin{tabular}{@{}c@{}}
        \textbf{Bf}
    \end{tabular}}
    & \textbf{\textit{SI-SDRi}}&\textbf{\textit{PESQ}}& \textbf{\textit{STOI}}& 
    \textbf{\textit{SI-SDRi}}&\textbf{\textit{PESQ}}& \textbf{\textit{STOI}}& 
    \textbf{\textit{SI-SDRi}}&\textbf{\textit{PESQ}}& \textbf{\textit{STOI}}&  
    \textbf{\textit{SI-SDRi}}&\textbf{\textit{PESQ}}& \textbf{\textit{STOI}}\\
     &  &  &\textpm0.07 & \textpm0.00 & \textpm0.00 & \textpm0.11 & \textpm0.01 & \textpm0.00 & \textpm0.08 & \textpm0.01 & \textpm0.00 & \textpm0.07 & \textpm0.00 & \textpm0.00\\ 
    \midrule
    Librispeech & Supervised & No & 6.19 & 1.37 & 0.58 & 5.09 & 1.32 & 0.55 & 6.38 & 1.37 & 0.58 & \textbf{11.88} & \textbf{1.89} & \textbf{0.79} \\ 
    and FSD50K &            & Yes & 9.89 & 1.53 & \textbf{0.62} & \textbf{8.66} &\textbf{1.50} & \textbf{0.61} & 11.72 & 1.58 & \textbf{0.64} & - & - & - \\ % run id oxf5w43t
                & Unsupervised & No & 4.21 & 1.22 & 0.54 & 3.20 & 1.20 & 0.52 & 4.27 & 1.23 & 0.54 & 9.68 & 1.60 & 0.74 \\
                &            & Yes & 8.37 & 1.41 & 0.59 & 7.07 & 1.39 & 0.58 & 11.12 & 1.61 & 0.63 & - & - & - \\ % run id 4do5uglh
                & Unsup. w/ Weighting & No & 4.21 & 1.23 & 0.55 & 3.07 & 1.20 & 0.52 & 4.08 & 1.22 & 0.54 & 9.54 & 1.59 & 0.74 \\
                &            & Yes & 8.25 & 1.41 & 0.60 & 6.87 & 1.37 & 0.59 & 10.66 & 1.57 & 0.63 & - & - & - \\ % run id yxwdeefy
    \midrule
    AudioSet & Unsupervised & No & 5.10 & 1.31 & 0.54 & 3.80 & 1.26 & 0.49 & 4.70 & 1.29 & 0.52 & 4.50 & 1.42 & 0.58 \\
             &            & Yes & 9.68 & \textbf{1.62} & 0.60 & 7.76 & \textbf{1.50} & 0.57 & 11.86 & \textbf{1.83} & 0.63 & - & - & - \\ %run id o5u1z7wa
             & Unsup. w/ Weighting & No & 5.30 & 1.30 & 0.54 & 3.80 & 1.23 & 0.48 & 4.88 & 1.27 & 0.52 & 4.69 & 1.42 & 0.58 \\
             &            & Yes & \textbf{9.95} & 1.60 & 0.60 & 7.67 & 1.45 & 0.55 & \textbf{12.28} & 1.81 & 0.63 & - & - & - \\ %run id ng1nh4wp
    \bottomrule
  \end{tabular}
\end{table*}

\endgroup

Tables \ref{tab:results_bark} and \ref{tab:results_pesq_stoi} present the results of different training methods on 4 test datasets. The 3 from MCFSTD (ReSpeaker, Kinect and 16Sounds) and the final one is the test dataset from FSD50K or Librispeech (Single) \cite{panayotov_librispeech_2015}. To create the Single test dataset and supervised training datasets, isolated targets are necessary. For target signals in bark enhancement, we use audio samples in FSD50K that only has the bark label as target. For speech, we use samples from Librispeech augmented with RIRs from BIRD \cite{grondin2020bird}. For the non-target signals in both cases, we use the samples in FSD50K that does not include the target class. To create MoMs, we iterate 200 times on the Single bark target mixtures, mixing them with different interference. This is done 10 times for each other dataset. We do not iterate for target only results. For testing and training, the MoM contains 2 to 4 mixtures. The gain of every mixture is normalized randomly to between -5 and 5 dB to add more robustness in the network. In both speech and bark enhancement, there is a noticeable improvement across all cases when beamforming is applied, as opposed to relying solely on the network predictions. The weighting leads to slight improvements in some cases.

We observe a difference in the supervised results compared to the unsupervised results for speech enhancement as opposed to bark enhancement. This is due to the amount of clean isolated data available for supervised training. In our supervised training datasets, there are 100.6 hours of speech samples and only 0.4 hours of bark samples. In the case of bark enhancement, unsupervised training performs better than supervised training in and out of domain. In-domain test data is defined as audio recorded in the same conditions as samples used for training. While for speech enhancement, supervised training proves more effective in domain and, in certain instances, for out-of-domain datasets. This highlights the benefits of unsupervised training when there is not a readily amount of clean data available for a particular target sound. This also diminishes the workload required for gathering the data essential to train a neural network.

The robustness in change of domain with unsupervised training compared to supervised training can also be observed in the results as first noted in \cite{wisdom_unsupervised_2020}. It is possible to observe this by subtracting the SI-SDRi of the in-domain test dataset (Single) with the SI-SDRi of the out-of-domain datasets (MCFSTD). On average, we note a larger performance drop of 0.31 dB for supervised training in contrast to unsupervised training. However, this is not possible to observe for models trained on AudioSet as the Single dataset is also out-of-domain. Domain robustness is crucial for training a neural network that will maintain high performance once deployed in real-world scenarios, 
particularly when using training data from the exact target domain is impossible. 

When looking at the results for speech enhancement, it is also possible to observe that the supervised network outperforms the unsupervised networks trained on AudioSet \cite{gemmeke_audio_2017}. However, once the beamforming is applied, the inverse is observed in the SI-SDRi for the ReSpeaker and 16Sounds and in the PESQ for the MCFSTD datasets. We hypothesize that this is because the unsupervised networks seems to be less aggressive in suppressing interference than the supervised networks. This results in a reduced presence of the target in the noise SCM but an increased amount of interference in the target SCM compared to the supervised network prediction. This seems to help the MVDR beamformer as its objective is to minimize the power of the noise while constraining the distortion in the target direction \cite{souden_optimal_2010}. Having a smoother noise SCM can also contribute to a better numerical stability when computing the inverse of the noise SCM. This is important when using a mask predicted by a neural network because the SCM can be very sparse in some frequencies, as first noted in \cite{heymann_neural_2016}.

\section{Conclusions}

UIMVDR enables multi-channel sound enhancement or separation to benefit from large weakly labelled or unlabelled datasets. The SI-SDRi, PESQ and STOI showed the advantages of using an unsupervised single channel neural network with an MVDR beamformer to improve estimation in real conditions. Especially for sounds where it is difficult to collect clean isolated data in the domain of the intended use. The results were obtained using a new test dataset, the MCFSTD.

\section{Acknowledgements}

The work reported here was supported by the Natural Sciences and Engineering Research Council of Canada (NSERC) and by the Fonds de Recherche du Québec en Nature et Technologies (FRQNT). We would also like to express our gratitude to Charles Maheu for his help with the experimental setup.

\bibliographystyle{IEEEtran}

\bibliography{mybib}

\end{document}